\documentclass[12pt]{article}
\usepackage{amsmath,amssymb}
\usepackage{graphicx,epsfig}
\newcommand{\abs}[1]{\left\vert#1\right\vert}
\newcommand{\gsim}{\, \mathop{}_{\textstyle \sim}^{\textstyle >} \,}
\newcommand{\lsim}{\, \mathop{}_{\textstyle \sim}^{\textstyle <} \,}
\def\spur#1{\mathord{\not\mathrel{#1}}}

\begin{document}
\baselineskip 0.6cm

\begin{titlepage}

\begin{flushright}
UCB-PTH-02/36 \\
LBNL-51300 \\
\end{flushright}

\vskip 1.0cm

\begin{center}
{\Large \bf Radiative Electroweak Symmetry Breaking  \\
  from a Quasi-Localized Top Quark}

\vskip 1.0cm

{\large
Riccardo Barbieri$^a$, Lawrence J.~Hall$^{b,c}$, 
Guido Marandella$^a$, Yasunori Nomura$^{b,c}$, 
Takemichi Okui$^{b,c}$, Steven J.~Oliver$^{b,c}$ and 
Michele Papucci$^a$}

\vskip 0.5cm

$^a$ {\it Scuola Normale Superiore and INFN, Piazza dei Cavalieri 7,
                 I-56126 Pisa, Italy} \\
$^b$ {\it Department of Physics, University of California,
                Berkeley, CA 94720, USA}\\
$^c$ {\it Theoretical Physics Group, Lawrence Berkeley National Laboratory,
                Berkeley, CA 94720, USA}

\vskip 1.0cm

\abstract{We consider 5D supersymmetric $SU(3) \times SU(2) \times U(1)$ 
theories compactified at the TeV scale on $S^1/Z_2$ with supersymmetry 
broken by boundary conditions. Localizing the top quark at a boundary 
of a fifth dimension by a bulk mass term $M_t$, reduces the strength of 
radiative electroweak symmetry breaking. For $M_t R \approx 1$--$2$, 
the natural value for the top and bottom squark masses are raised to 
$500$--$1200~{\rm GeV}$, and all other superpartners may have masses 
of the compactification scale, which has a natural range of $1/R \simeq 
1.5$--$3.5~{\rm TeV}$. The superpartner masses depend only on $1/R$, and 
are precisely correlated amongst themselves and with the mass of the 
Higgs boson, which is lighter than $130~{\rm GeV}$.}

\end{center}
\end{titlepage}

\section{Introduction}
\label{sec:intro}

While the past decade has been characterized by the success of the
precision tests of the Standard Model (SM), the present decade, with
the progression of the Tevatron runs and especially with the coming
into operation of the LHC, should allow a thorough exploration of the
physics of ElectroWeak Symmetry Breaking (EWSB), which remains a central 
unsettled problem in  the theory of the fundamental interactions.

To say that EWSB is an unsettled problem does not quite do justice
to the Standard Model. Although not directly proven experimentally, the
SM most likely captures the essence of EWSB through the Higgs mechanism.
The limitation of the SM in the EWSB sector is rather the lack of
quantitative predictive power: the two parameters of the Higgs
potential, $\mu$ and $\lambda$, are in one to one correspondence with the
two physical observables, the Higgs mass and the Higgs vacuum expectation 
value (VEV), or the Fermi constant.

To improve on this situation is non-trivial, to say the least, due to 
the lack of crucial data so far. Among the different attempts, the one 
that goes farther involves supersymmetry, whose breaking triggers EWSB.
Other than being supported by the success of gauge coupling unification,
the standard supersymmetric picture of EWSB improves also on the 
predictive power of the SM. The quartic coupling $\lambda$ gets related 
to the gauge couplings, up to significant but controllable radiative
corrections, thus implying a light (too light?) Higgs boson. Similarly 
the Fermi scale is given in terms of the soft supersymmetry-breaking
parameters, $m_i$, in turn related to the superpartner masses, with only 
mild (logarithmic) dependence on the cutoff scale $\Lambda$,
\begin{equation}
  G_F = G_F(m_i, \log\Lambda).
\end{equation}

The explicit form of this equation, although somewhat model dependent,
is also the basis for expecting superpartners near the Fermi scale. 
Yet its structure and the number of parameters it generally involves 
has not made possible, so far, any precise statement on the superpartner 
masses: all quantitative predictions for them rest upon forbidding some 
predetermined level of fine-tuning among the different parameters. 
Although this is a plausible attitude to take, it does not avoid the 
most unpleasant feature of the current supersymmetric extensions of 
the SM. How should one judge, for example, in a truly objective way the 
significance of the failure to find any superpartner (or the Higgs 
itself) at LEP?  Similarly, or conversely, how should one determine in 
a precise manner the discovery potential of supersymmetry at the 
Tevatron or even at the LHC?

This difficulty could be due to the inadequacy of the present
understanding of supersymmetry breaking. For this reason and with the
aim at drastically reducing the number of relevant parameters, a
concrete model of supersymmetry breaking has been proposed in 
Ref.~\cite{Barbieri:2000vh}, whose basic content is to establish 
a precise connection between the Fermi scale and the inverse radius 
of a compactified fifth dimension $1/R$.

That this is possible at all looks at first rather surprising in view
of the non-renormalizability of field theories in 5D. In this respect 
there are two crucial properties of this model: 
i) the existence of a residual local supersymmetry (and of a global 
$U(1)_R$ symmetry) that highly restrict the form of the Lagrangian 
with its possible counterterms;
ii) the description of EWSB {\it and} of all fermion masses in terms 
of a single Higgs doublet as in the SM and unlike the case of the 
Minimal Supersymmetric Standard Model (MSSM). The cost of 
non-renormalizability of the 5D Lagrangian is only the requirement
of a rather low cutoff scale $\Lambda$. This is not, however, a serious 
limitation in so far as one can show that the Higgs physics and the 
physics at $1/R$ have only weak ultraviolet sensitivity in an effective 
field theory sense.

One could object at this point that a low cutoff obscures the
motivation coming from the gauge coupling unification and, therefore,
any motivation at all for the entire program. Although we agree that
the apparent lack of gauge unification in the model of 
Ref.~\cite{Barbieri:2000vh} is a step backward, we think that 
achieving a description of EWSB which involves a naturally light Higgs 
perturbatively interacting up to a multi-TeV cutoff scale $\Lambda$ 
is both non-trivial and quite clearly motivated by the current status 
of the ElectroWeak Precision Tests (EWPT).\footnote{
The low Confidence Level of the current SM fit of the EWPT has raised 
some questions on this last statement \cite{Chanowitz:2001bv, 
Novikov:2002tk}. We find the causes of the poor fit not particularly
troublesome and, henceforth, the evidence for the lightness of the
Higgs boson significant, although indirect and as such subject to
obvious limitations. We thank Alessandro Strumia for help in
clarifying this issue.}

The theories presented in this paper have the same motivation as the 
model of Ref.~\cite{Barbieri:2000vh} and share with it the key 
properties that make possible a quantitative connection between $1/R$ 
and $G_F$.  We study top quark hypermultiplets with bulk mass terms, 
so that the top quark wavefunctions are peaked close to a boundary of 
the fifth dimension, rather than being smoothly distributed
throughout the bulk.  This quasi-localization also yields controlled 
EWSB, but with the important result that the compactification scale is 
significantly larger, $1/R \simeq 1.5$--$3.5~{\rm TeV}$, than in the 
Constrained Standard Model (CSM) of Ref.~\cite{Barbieri:2000vh}, giving 
a more natural agreement with EWPT.  The cutoff scale is also increased 
and can be as large as $15~{\rm TeV}$ in the entire range of $1/R$.

\section{$U(1)_R$ Invariant Theories}
\label{sec:theory}

A predictive theory of EWSB should have only a few parameters, and 
therefore as much symmetry as possible. We study theories of a single 
extra dimension, with gauge group $SU(3) \times SU(2) \times U(1)$, 
compactified to a line segment $(0, \pi R/2)$.  We consider the 
possibility that the bulk Lagrangian has the following symmetries:
\begin{itemize}
\item 5D supersymmetry. The resulting bulk gauge interactions possess 
 a $SU(2)_R$ symmetry. The physical line segment in the fifth dimension 
 may be viewed as arising from an orbifold compactification corresponding 
 to two orbifold symmetries: a translation by $\pi R$ involving 
 a rotation angle $\alpha$ inside $SU(2)_R$, and a reflection parity 
 about a particular point, which we label $y=0$ \cite{Barbieri:2001dm}.
 For the case of interest to us $\alpha \neq 0$, so that the 
 compactification leaves no unbroken supersymmetry in the low energy 
 equivalent 4D theory but a residual local supersymmetry in 5D. 

\item An additional unbroken symmetry, $U(1)_R$, with quantum numbers 
 shown in Table~\ref{R-charge}. It arises from the special case of 
 $\alpha = 1/2$, which  corresponds to the $S^1/(Z_2 \times Z'_2)$ 
 orbifold compactification introduced in Ref.~\cite{Barbieri:2000vh}. 
 This symmetry ensures that Majorana gaugino masses, $A$-terms or the 
 $\mu$ of the Higgs sector are not generated from either 
 compactification or brane interactions. 
\begin{table}
\begin{center}
\begin{tabular}{|c||c|c|c|} \hline
$R$  & gauge $V$       & Higgs $H$     & matter $M$               \\ \hline
+2   &                 & $h^c$         &                          \\ 
+1   & $\lambda$       & $\tilde{h}^c$ & $\tilde{m}, \tilde{m}^c$ \\ 
0    & $A_\mu, \Sigma$ & $h$           & $m, m^c$                 \\ 
$-1$ & $\lambda'$      & $\tilde{h}$   &                          \\ \hline
\end{tabular}
\caption{Continuous $R$ charges for gauge, Higgs and matter multiplets.
Here, $A_\mu$, $(\lambda, \lambda')$ and $\Sigma = (\sigma + iA_5)$ are 
the gauge field, two gauginos and the adjoint scalar inside a 5D gauge 
multiplet; $(h, h^c)$ and $(\tilde{h}, \tilde{h}^c)$ are two complex 
scalars and two Weyl fermions inside a 5D Higgs hypermultiplet; and 
$(m, m^c)$ and $(\tilde{m}, \tilde{m}^c)$ are two complex scalars 
and two Weyl fermions inside a 5D matter hypermultiplet, where $m$ 
represents $q, u, d, l$ and $e$.}
\label{R-charge}
\end{center}
\end{table}

\item A local parity $P_5$, corresponding to a reflection about any 
 point of the bulk (in the limit $R \rightarrow \infty$). This 
 symmetry forbids both a Chern-Simons term and bulk masses for the 
 hypermultiplets. However, with only one Higgs hypermultiplet containing 
 a massless scalar $H_u$, as in Ref.~\cite{Barbieri:2000vh}, consistency 
 of the theory requires the breaking of this symmetry 
 \cite{Arkani-Hamed:2001is, Barbieri:2002ic}, while introducing 
 a quadratically divergent brane-localized Fayet-Iliopoulos (FI) 
 term \cite{Ghilencea:2001bw, Barbieri:2001cz}. Hence, to consider 
 this symmetry we must introduce a second Higgs hypermultiplet with 
 boundary conditions that give rise to a second massless scalar $H_d$ 
 with opposite hypercharge to $H_u$.\footnote{
 Here $H_u$ and $H_d$ do not necessarily correspond to the fields 
 giving up-type and down-type quark masses, respectively: for instance, 
 down-type quark masses can arise from the VEV of $H_u$, as seen in 
 Eq.~(\ref{eq:Yukawas}).}
\end{itemize}

Is it possible to construct a completely realistic theory with these 
three symmetries? Other than gauge interactions, the symmetries allow 
brane-localized Yukawa interactions. For the case of bulk matter 
these are:
\begin{eqnarray}
  {\cal L}_{\rm Yukawa} &=& 
     \delta(y) [\lambda_u Q U H_u + \lambda_d Q D H_d]_{\theta^2} 
\nonumber\\
  && +\delta(y-\pi R/2) [\lambda'_u Q' U' H_d^{\prime c} 
     + \lambda'_d Q' D' H_u^{\prime c}]_{\theta^{\prime 2}},
\label{eq:Yukawas}
\end{eqnarray}
where $Q,U,D$ and $H_{u,d}$ are chiral multiplets containing quark 
and Higgs-boson zero modes of the $N=1$ supersymmetry acting at $y=0$, 
while $Q',U',D'$ and $H^{\prime c}_{u,d}$, which also contain quark 
and Higgs-boson zero modes, are the chiral multiplets of the $N=1$ 
supersymmetry acting at $y = \pi R/2$.  Even though one-loop radiative 
corrections lead to contributions to the soft mass terms $m_u^2 
H_u^\dagger H_u + m_d^2 H_d^\dagger H_d + m_3^2(H_u H_d + {\rm h.c.})$ 
in the Higgs potential, successful EWSB does not occur. The Yukawa 
contributions dominate $m_{u,d}^2$ and are large and negative, so that 
$m_u^2 + m_d^2 < 0$, giving an unbounded potential along the $D$-flat 
direction.  It is interesting that the addition of an extra Higgs 
doublet hypermultiplet to the CSM destroys the theory. If the quark 
fields reside on a boundary, only one pair of the Yukawa couplings 
survive \cite{Pomarol:1998sd}. In this case the squark masses arise 
only at one loop, and we find that the corresponding two-loop top 
Yukawa contribution to $m_u^2$ is not sufficiently negative to overcome 
the positive contribution from the one-loop gauge radiative correction: 
$m_{u,d}^2$ are both positive, and $m_3^2 = 0$, so that there is no EWSB.

We conclude that we must give up either the bulk parity $P_5$ or the 
continuous $U(1)_R$ symmetry to construct realistic theories. 
Theories with $P_5$ but no $U(1)_R$ were constructed in 
Ref.~\cite{Arkani-Hamed:2001mi}. They are theories with two Higgs 
doublet VEVs resulting from a scalar potential having terms induced 
by $U(1)_R$ breaking boundary operators. Here we pursue the alternative 
possibility of a $U(1)_R$ symmetric theory with $P_5$ broken.  
Once $P_5$ is given up, we can introduce bulk mass terms for the 
hypermultiplets \cite{Barbieri:2002ic}: ${\cal L} = [M_\Phi \Phi 
\Phi^c]_{\theta^2}$.  Since there is no $P_5$ symmetry, we study 
both one Higgs and two Higgs hypermultiplet versions of the theory. 
Note that the one Higgs version of the theory is precisely the theory 
introduced in Ref.~\cite{Barbieri:2000vh} with the hypermultiplets 
having non-vanishing bulk masses.  While the one Higgs theories have 
a quadratically divergent FI term, the two Higgs theories are less 
sensitive to unknown physics at the cutoff, as we discuss shortly.

From the viewpoint of EWSB the bulk mass terms of most importance 
are those of the third generation and the Higgs multiplets. For most 
of this paper we assume that the bulk mass for the $H_u$ hypermultiplet 
vanishes $M_{H_u} = 0$, and we concentrate on the bulk masses for the 
third generation quarks: $M_{Q,U,D}$. We consider values of $M_{Q,U}$ 
comparable to or larger than $1/R$, so that the corresponding zero-mode 
wavefunctions are peaked at the boundaries of the fifth dimension. 
In particular we choose $M_Q$ positive\footnote{
Note that our sign convension is the opposite of the one used 
in \cite{Barbieri:2002uk}.
} 
so that the left-handed top 
and bottom quarks are located near $y=0$. To avoid large wavefunction 
suppression factors in the top quark mass we also choose $M_U$ to be 
positive. There is still freedom in $M_D$, which we allow to be positive, 
negative or even zero. In the theory with a single Higgs hypermultiplet, 
the $b$ quark must get its mass from a Yukawa coupling at $y= \pi R/2$, 
so that no matter which choice is made for $M_D$, the $m_b/m_t$ mass 
ratio receives a suppression of at least $\exp(-\pi M_Q R/2)$ due to the 
small value of the $q$ wavefunction at $y= \pi R/2$ \cite{Marti:2002ar}. 
It is significant that localization necessarily destroys the symmetry 
between up and down sectors, leading to a small value for $m_b/m_t$ 
without the need for a hierarchy of 5D Yukawa couplings.

In the theory with two Higgs hypermultiplets the situation is 
complicated by the possibility of four Yukawa couplings, as shown in 
Eq.~(\ref{eq:Yukawas}), and two Higgs VEVs. Nevertheless, we require that 
localization naturally yield a hierarchy for $m_b/m_t$ and find that this 
can emerge in two ways:
\begin{itemize}
\item Introduce a global symmetry $U(1)_{H_d}$ which rotates the phase 
 of only $H_d$, and therefore sets the Yukawa couplings $\lambda'_u = 
 \lambda_d = 0$ as well as $m_3^2 = 0$. Only $H_u$ acquires a VEV, so 
 that the physics of both EWSB and quark mass generation is identical 
 to the case of the one Higgs theory.

\item Break the symmetry between up and down sectors by requiring that 
 $D$ not be localized at $y=0$. In this case, even though all Yukawa 
 couplings may be comparable, the radiatively generated value for 
 $m_3^2$ is very small so that the VEV of $H_d$ is negligible, and 
 the $t$ and $b$ quark masses arise dominantly from $\lambda_u$ and 
 $\lambda'_d$. Furthermore the contributions of the other two Yukawa 
 couplings to $m_{u,d}^2$ are also negligible. 
\end{itemize}

In all these theories, the Higgs potential, and therefore EWSB, depends 
only on the unknown parameters $1/R, M_Q, M_U$ and $M_{H_u}$. The absence 
of any dependence on other bulk mass parameters, in particular $M_D$, is 
discussed in Appendix A. The top Yukawa coupling $\lambda_u$ enters, but 
is determined by $m_t$. In the next section we study the region of 
parameter space with $M_Q = M_U$ and $M_{H_u} = 0$ and find a restricted 
and therefore predictive region in which EWSB is successful. In section 
\ref{sec:ewsb2} we study the dependence on $M_U/M_Q$ and small values of 
$M_{H_u}$, and find that successful EWSB persists.  The calculation of 
EWSB is identical for the one Higgs theory and for both types of two 
Higgs theories. The origin of EWSB is always a radiative contribution 
to $m_u^2$ from a top quark hypermultiplet which has a wavefunction 
peaked around $y=0$.

Finally, we discuss the FI term. Since $P_5$ is broken, a brane-localized
FI term for the hypercharge gauge interaction is allowed in our theories
at tree level and can be generated by radiative corrections. However, 
by shifting the VEV of the scalar in the hypercharge gauge multiplet, 
we can always transform the FI term to bulk masses for the hypermultiplets 
\cite{Barbieri:2002ic}.  Therefore, the FI term does not represent an 
additional parameter, as long as we consider all the bulk hypermultiplet 
masses in the analysis. The hypermultiplet masses in this paper are 
meant to be the ones after this transformation: they include both the 
tree and radiative contributions of the FI term.  Then, even when the 
FI term is quadratically divergent, as in the case of the one Higgs 
theory, the resulting hypermultiplet masses are small relative to $1/R$. 
In the case $M_{Q,U} \ll 1/R$, the induced value for $M_{H_u}$ produces 
only a small perturbation to the theory \cite{Barbieri:2002uk}.
However, in the case of a quasi-localized top quark, $M_{Q,U} \gsim 
1/R$, the induced value for $M_{H_u}$ gives a mass squared to the 
lightest mode of $H_u$ of comparable size to the other radiative 
contributions.  The quadratic divergence of the FI term can be canceled 
by a second Higgs hypermultiplet --- indeed this may be a motivation for 
considering two Higgs theories.  In the presence of hypermultiplet 
masses, however, a further condition arises from the ultraviolet 
insensitivity of the FI term, since there is a residual linear divergence 
proportional to $\mathrm{Tr}[YM]$, where $Y$ and $M$ are hypercharge 
and bulk-mass matrices for the hypermultiplets. This may motivate
interesting relations among the hypermultiplet masses, e.g., the case 
in which they are all equal, or $M_Q=M_U=M_D$ or $M_Q=M_U= -M_{H_d}$ 
and all other masses equal to zero. These relations make the radiative 
FI term identically vanishing.  In fact, $M_{H_u}=0$ then becomes 
a perfectly stable condition.

\section{Electroweak Symmetry Breaking}
\label{sec:ewsb1}

In this section we study in detail the EWSB when $M_Q=M_U \equiv M_t$
and $M_{H_u}=0$.  We are mainly interested in the region where $M_t R 
\gsim 1$.  The tree-level potential for the Higgs is
\begin{equation}
  V_{\textrm{tree}}(\phi) = \frac{g^2+g^{\prime 2}}{8}\abs{\phi}^4
    = \frac{m^2_Z}{4 v^2}\abs{\phi}^4,
\end{equation}
where $\phi$ is the neutral component of $H_u$. EWSB is triggered by 
radiative corrections, which requires studying the corrected effective 
potential $V(\phi)$.  One important point of our analysis is that a 
(quasi-)localized top quark naturally gives a large $1/R$ compared to 
the weak scale, as we show in subsection \ref{subsec:quad}. The full 
detail of the effective potential is presented in subsection 
\ref{subsec:full}.

\subsection{Localized matter and large $1/R$}
\label{subsec:quad} 

\begin{figure}
\begin{center}
  \includegraphics[width=10cm]{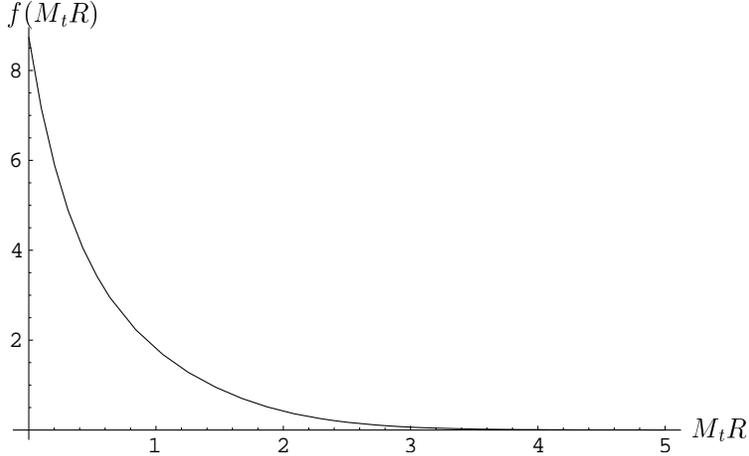}
\caption{The function $f(M_t R)$ appearing in the one-loop top 
 contribution, Eq.~(\ref{eq:1looptopquad}).  The exponential drop 
 off of $f(M_t R)$ at large $M_t R$ is due to the recovery of a 
 supersymmetric spectrum of the top-stop towers in this limit, 
 in particular, due to the recovery of the supersymmetric mass 
 degeneracy for the chiral multiplets made of the lightest 
 modes, which get localized at $y=0$ and become massless 
 \cite{Barbieri:2002uk}.}
\label{fig:f-mtr}
\end{center}
\end{figure}

To demonstrate a couple of important points in our analysis, it is best 
to discuss separately the corrections to the quadratic term,
\begin{equation}
  V^{(2)}(\phi) = 
    V^{(2)}_{\textrm{1loop}}(\phi) + V^{(2)}_{\textrm{2loop}}(\phi),
\label{eq:pot-complete}
\end{equation}
from the rest of the potential, $\delta V(\phi)$: $V(\phi) = V^{(2)}(\phi) 
+ \delta V(\phi)$. In the one-loop quadratic piece 
\begin{equation}
  V^{(2)}_{\textrm{1loop}}(\phi) 
    \equiv V^{(2)}_{\textrm{1loop,gauge}}(\phi) 
    + V^{(2)}_{\textrm{1loop,top}}(\phi),
\label{eq:pot-1loop}
\end{equation}
we include the $SU(2)$ and $U(1)$ gauge contributions to the Higgs 
squared mass \cite{Antoniadis:1998sd}
\begin{equation}
  V^{(2)}_{\textrm{1loop,gauge}}(\phi)
    = \frac{A_{SU(2)}+A_{U(1)}}{R^2} \abs{\phi}^2 
    = 0.76 \frac{0.01}{R^2} \abs{\phi}^2,
\label{eq:gauge1loophiggs}
\end{equation}
where
\begin{eqnarray}
  && A_{SU(2)} 
   = \frac{21 \zeta(3) g^2}{16 \pi^4}
   = 0.00688,
\label{eq:gauge1loophiggssu2}\\
  && A_{U(1)} 
   = \frac{7 \zeta(3) g^{\prime 2}}{16 \pi^4}
   = 0.00069,
\label{eq:gauge1loophiggsu1}
\end{eqnarray}
and the one-loop top-stop corrections at arbitrary $M_t R$, as
computed in Ref.~\cite{Barbieri:2002uk}, which is given by
\begin{equation}
  V_{\textrm{1loop,top}}^{(2)}(\phi) 
    = - f(M_t R) \frac{0.01}{R^2} \abs{\phi}^2,
\label{eq:1looptopquad}
\end{equation}
where $f(M_t R)$ is plotted in Fig.~\ref{fig:f-mtr}.

The masses of the scalar components of the lightest modes (the 
left-handed and right-handed stops) also receive a significant
correction at one loop, due to the gauge and top-Yukawa interactions. 
These corrections have been computed in Ref.~\cite{Delgado:1998qr} 
in the case of exact localization of the chiral top multiplets with 
the result
\begin{eqnarray}
  && m^2_{\widetilde{Q}} 
    = \frac{28 \zeta (3)}{3 \pi^3} \frac{\alpha_s}{R^2} 
      + \frac{7 \zeta (3)}{2 \pi^3} \frac{\alpha_t}{R^2} 
    =  \frac{0.052}{R^2}, 
\label{eq:1loopmassQ}\\
  && m^2_{\widetilde{U}} 
    = \frac{28 \zeta (3)}{3 \pi^3} \frac{\alpha_s}{R^2} 
      + 2 \frac{7 \zeta (3)}{2 \pi^3} \frac{\alpha_t}{R^2} 
    = \frac{0.062}{R^2}, 
\label{eq:1loopmassU}
\end{eqnarray}
where $\alpha_s \equiv g_3^2/4\pi$ and $\alpha_t \equiv y_t^2/4\pi$ 
with $g_3$ and $y_t$ representing the QCD and top-Yukawa couplings in 4D.
The error from using exactly localized matter is expected to be small;
in fact we have checked that the $\alpha_t$ piece above deviates from 
the exact result by less then $15\%$ at $M_t R \gsim 1$.

\begin{figure}
\begin{center}
  \includegraphics{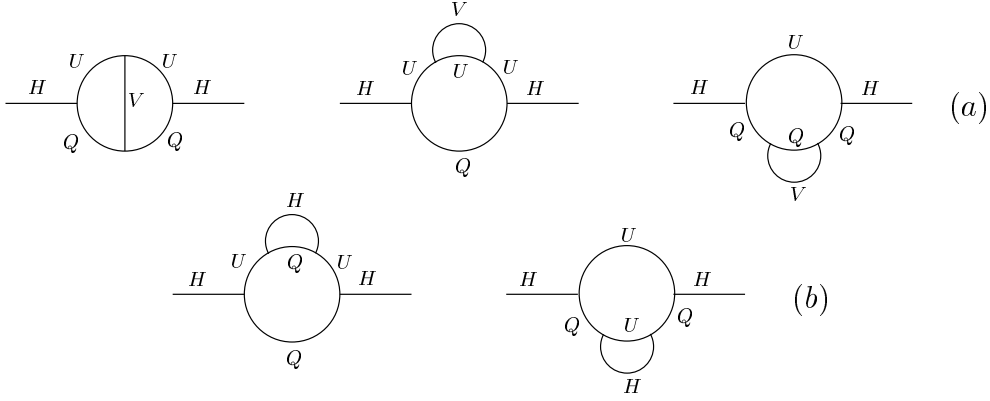} 
\caption{The diagrams contributing to the Higgs mass squared 
 at order (a) $\alpha_s \alpha_t$ and (b) $\alpha_t^2$.}
\label{fig:2loopdiag}
\end{center}
\end{figure}

The vanishing of $V_{\textrm{1loop,top}}^{(2)}(\phi)$ as $M_t R$
increases makes it necessary to compute the dominant two-loop
effects. The diagrams that contribute to the Higgs mass squared 
at order $\alpha_s \alpha_t$ and $\alpha_t^2$ are shown in
Fig.~\ref{fig:2loopdiag}(a) and \ref{fig:2loopdiag}(b), respectively, 
in superfield notation.
We compute them with localized $Q$ and $U$ chiral multiplets. 
Note that these diagrams have to be ultraviolet finite without any 
subtraction because in localized approximation for $Q$ and $U$ the 
Higgs squared mass has no $\alpha_t$ (nor $\alpha_s$) contribution 
which would generate a two-loop counterterm by renormalizing 
$\alpha_t$ (or $\alpha_s$).

The calculation can be done by means of the propagators for 
the various components of the gauge and Higgs supermultiplets in 
mixed (4 momentum)-(5th coordinate) space, $G_i(k_4;y,y')$. 
Specifically, in localized approximation for $Q$ and $U$, one needs 
these propagators at $y=y'=0$ given in Ref.~\cite{Barbieri:2002uk}.
Summing up the various contributions from the diagrams in 
Fig.~\ref{fig:2loopdiag}, expressed in components and without 
eliminating the auxiliary fields, one finds
\begin{equation}
  V_{\textrm{2loop}}^{(2)}(\phi) 
    = -\abs{\phi}^2 \left(144 \pi^3 R \alpha_t^2 
      + 256 \pi^3 R \alpha_t \alpha_s \right) 
      \int \frac{\textrm{d}^4 p \; \textrm{d}^4 q}{(2 \pi)^8} 
      \frac{q}{p^4 (p-q)^2} \left(\coth \frac{\pi R q}{2} 
      - \tanh \frac{\pi R q}{2} \right),
\label{eq:2looppot}
\end{equation}
where $p$ and $q$ are Euclidean 4 momenta.

As expected, Eq.~(\ref{eq:2looppot}) exhibits the characteristic 
exponential convergence of the integrand at high momenta relative 
to $1/R$. On the contrary there is an infrared divergence due to 
the masslessness of the stops, which we cut off by giving the 
masses given in Eqs.~(\ref{eq:1loopmassQ}, \ref{eq:1loopmassU}) 
for the internal stop propagators. The overall result is
\begin{equation}
  V_{\textrm{2loop}}^{(2)}(\phi)
    = -\frac{\abs{\phi}^2}{R^2} \left[ \frac{3 \alpha_t^2}{8 \pi} 
      \left(2\eta(m_{\widetilde{U}}R) + \eta(m_{\widetilde{Q}}R)\right) 
    + \frac{8 \alpha_t \alpha_s}{8\pi} \left(\eta(m_{\widetilde{U}}R) 
      + \eta(m_{\widetilde{Q}}R)\right) \right],
\end{equation}
where
\begin{equation}
  \eta(y) = \int_0^{\infty} \textrm{d}x \; x^2 
    \log \left( 1+\frac{x^2}{y^2} \right) 
    \left( \coth\frac{\pi x}{2} - \tanh\frac{\pi x}{2} \right),
\end{equation}
which amounts to a negative contribution to the Higgs mass squared
\begin{equation}
  V_{\textrm{2loop}}^{(2)}(\phi) = -0.49 \frac{0.01}{R^2} \abs{\phi}^2.
\end{equation}

Note that the sum of $V^{(2)}_{\textrm{1loop,gauge}}(\phi)$ and 
$V_{\textrm{2loop}}^{(2)}(\phi)$, which is the only contribution 
for exactly localized matter, is {\it positive}.  This shows, as
anticipated in section \ref{sec:theory}, that no EWSB occurs in the 
case of matter localized on the boundary; theories with top quark 
located on the brane, such as those of Refs.~\cite{Pomarol:1998sd, 
Delgado:1998qr}, do not work.  To obtain a realistic theory, we need 
additional negative contributions to the Higgs mass squared. A simple 
possibility is to slightly delocalize the top quark from the brane: 
to use $V^{(2)}_{\textrm{1loop,top}}(\phi)$ to trigger EWSB.\footnote{
Another possibility is to completely delocalize the top quark, 
as in Refs.~\cite{Barbieri:2000vh, Barbieri:2002uk}, which leads to 
another interesting set of theories.}
Then, since the delocalization is not perfect, the contribution 
from $V^{(2)}_{\textrm{1loop,top}}(\phi)$ can still stay small and 
be comparable to those from $V^{(2)}_{\textrm{1loop,gauge}}(\phi)$ 
and $V_{\textrm{2loop}}^{(2)}(\phi)$, as explicitly shown in 
Eq.~(\ref{eq:1looptopquad}) and Fig.~\ref{fig:f-mtr}, naturally 
giving larger values for $1/R$ compared to the weak scale. 
This is one of the main results of our paper.  We study various 
consequences of this scenario in the rest of the paper.

\subsection{Full potential and EWSB} 
\label{subsec:full}

To complete the discussion of the EWSB, we need corrections to 
the remaining part of the potential:
\begin{equation}
  \delta V(\phi) \equiv 
    V_{\textrm{tree}}(\phi) 
    + \delta V_{\textrm{1loop}}(\phi) 
    + \delta V_{\textrm{2loop}}(\phi), 
\end{equation}
which are essential to obtain the physical Higgs-boson mass. 
At one loop we include the full top-stop contribution as a function 
of $M_t R$, which for large $1/R$ is given by 
\begin{eqnarray}
  && V_{\textrm{1loop,top}} (\phi) 
    = N_c \sum_{N=1}^{\infty} \int \frac{\textrm{d}^4 p}{(2\pi)^4} 
      \frac{(-1)^{N+1}}{N} 
      \left(\frac{y_t|\phi|}{\eta_0^q \, \eta_0^u}\right)^{2N} 
\nonumber \\
  && \left\{ \left[ G_{\varphi}^{U}(p;0,0) G_{F}^{Q}(p;0,0) \right]^N 
    + \left[ G_{\varphi}^{Q}(p;0,0) G_{F}^{U}(p;0,0) \right]^N 
    - 2 \left[ G_{\psi}^{U}(p;0,0) G_{\psi}^{Q}(p;0,0) \right]^N 
    \right\}, 
\label{eq:1looppot}
\end{eqnarray}
where $\eta_0^i$ is the wavefunction of the zero mode of particle $i$ 
evaluated at $y=0$; the forms of $G_i(p;0,0)$'s are given in Appendix B. 
At two loops we consider the correction
\begin{eqnarray}
  \delta V_{\textrm{2loop}}(\phi) 
  &=& \frac{3}{32 \pi^2} y_t^4 |\phi|^4 \Bigl[ 
    \log \bigl( y_t^2|\phi|^2 + m^2_{\textrm{tree}} 
      + m^2_{\widetilde{U}} \bigr) 
\nonumber\\
  && + \log \bigl( y_t^2|\phi|^2 + m^2_{\textrm{tree}} 
      + m^2_{\widetilde{Q}} \bigr) 
    - 2 \log \bigl( y_t^2|\phi|^2 + m^2_{\textrm{tree}} \bigr) 
    \Bigr],
\end{eqnarray}
where $m_{\textrm{tree}}(M_t R)$ is the common tree-level mass of 
the lightest stops at finite $M_t R$ and $m_{\tilde Q}^2$, 
$m_{\tilde U}^2$ are the radiative masses given in 
Eqs.~(\ref{eq:1loopmassQ}, \ref{eq:1loopmassU}). This is nothing but 
the top-stop radiative contribution to the quartic Higgs coupling in 
a standard MSSM effective potential with appropriate stop masses and 
no $A$-term, properly subtracted to avoid double countings with the 
one-loop top-stop corrections included in Eq.~(\ref{eq:pot-1loop}).

The two-loop mass term $V^{(2)}_{\textrm{2loop}}$ decreases slightly 
as $M_t R$ is brought down from infinity to $\sim 1$ to correctly 
induce EWSB.  In order to include all such dependences on the bulk mass, 
the actual calculations were performed with the full, properly subtracted, 
two-loop potential.  This necessitates cutting off the logarithm at some 
mass scale $M$ of order $1/R$, which is the scale where squark masses 
are generated.  The cutoff dependent terms in the potential approximate 
the contributions from higher Kaluza-Klein (KK) modes making the total 
potential finite. The precise value of $M$ was determined by matching 
the quadratic term with the exact calculation in the localized limit 
and found to be about $1.07/R$. (This cutoff itself may have some $M_t R$ 
dependence, but since the potential is only logarithmically sensitive we 
neglect any such correction.)
\begin{figure}
\begin{center}
  \includegraphics[width=15cm]{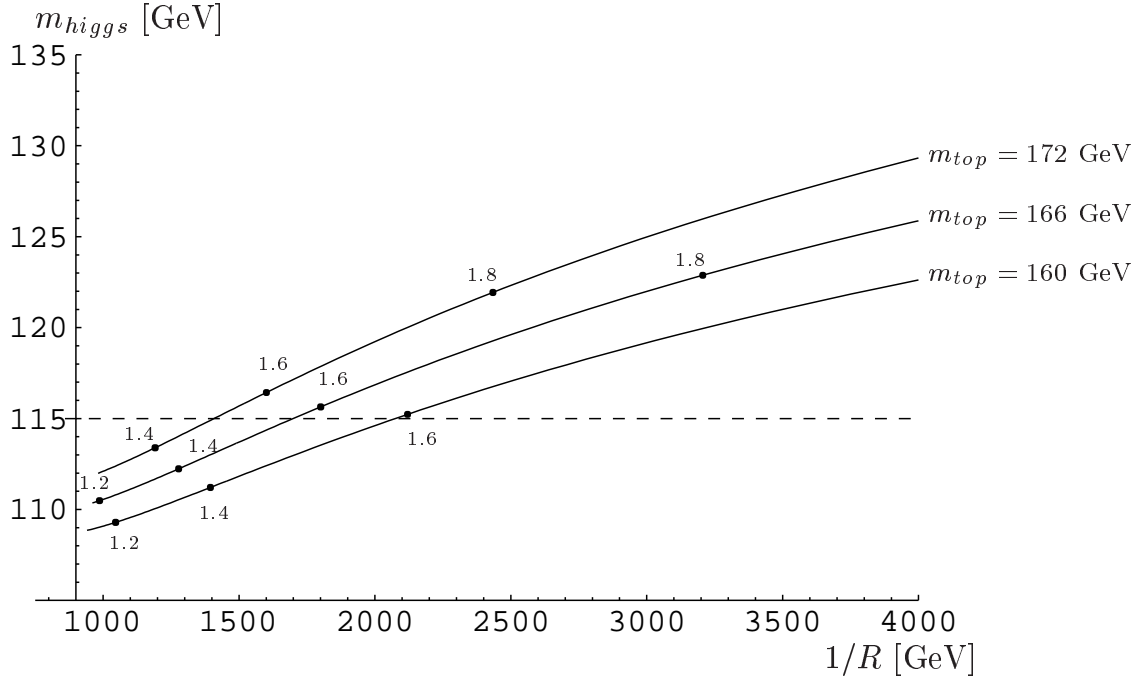}
\caption{The physical Higgs-boson mass as a function of $1/R$, obtained 
 by varying the top quark $\overline{MS}$ mass within the experimental
 $1 \sigma$ uncertainty.  The numbers written with dots on the curves 
 represent the values of $M_t R$ at the corresponding points.}
\label{fig:higgs-1overR}
\end{center}
\end{figure}

With the complete expression of the effective potential, which depends 
upon $1/R$ and $M_t$, we can now minimize it at $|\phi|=1/2 
( G_F/ \sqrt 2 )^{-1/2}$ and obtain a relation between these two 
parameters. As already mentioned, there is no EWSB in the localized 
approximation: the $|\phi|^2$ term in the potential is negative only 
for $M_t R \lsim 1.92$.  Note that if we were to (incorrectly) use the 
two-loop quadratic contribution in the localized limit, this value would 
rise to $M_t R \lsim 2.28$. If $H_d$ is present, on the contrary, 
for too low values of $M_t R$, the negative $m_u^2$-term in 
Eq.~(\ref{eq:1looptopquad}) may exceed the positive $m_d^2$-term and 
lead to an unstable potential, as discussed in section \ref{sec:theory}. 
The largest value of $M_t R$ at which this can happen is where no 
hypermultiplet mass for $H_d$ is introduced, i.e. at $M_t R \simeq 1.17$.

The resulting value of the physical Higgs-boson mass is shown in 
Fig.~\ref{fig:higgs-1overR} as a function of $1/R$ in the described 
range, for three different values of the running top quark mass, which 
determines $y_t$. This plot also shows the relevant range of $1/R$, 
which exceeds $1~{\rm TeV}$ at $M_t R \gsim 1.2$ because of the 
cancellation in the quadratic term of the effective potential 
already introduced. This cancellation becomes even stronger as 
$M_t R$ approaches 1.92 but turns into a fine-tuning region as 
$M_t R$ gets closer to the value of no EWSB.

\section{EWSB: Full Parameter Space}
\label{sec:ewsb2}

\begin{figure}
\begin{center}
  \includegraphics[width=9cm]{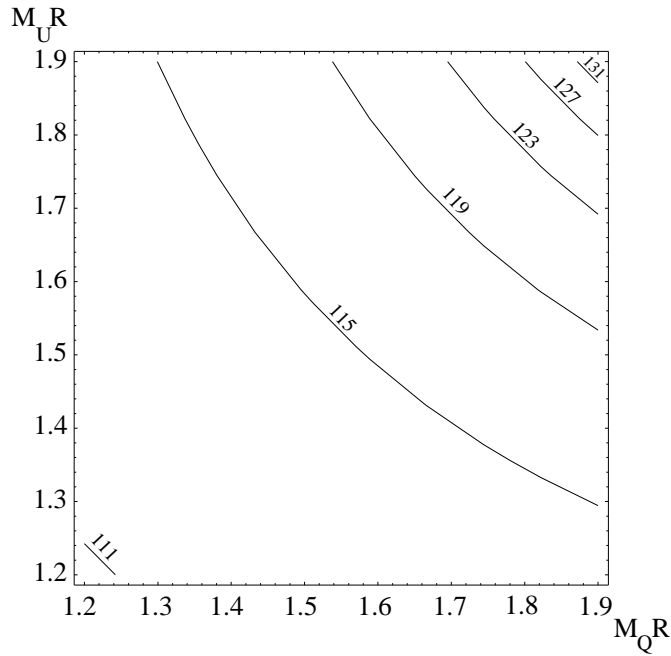}
\caption{Contours of the physical Higgs-boson mass in GeV as 
 a function of $M_{Q}R$ and $M_{U}R$.}
\label{fig:2DHiggs}
\end{center}
\end{figure}

We now consider the full parameter space of our model, namely 
$\left(M_{Q},M_{U},M_{H_{u}}\right)$.  Let us begin by removing the 
previous restriction $M_{Q}=M_{U}$.  This modifies the Higgs effective 
potential in a straightforward way.  The mixed momentum-position 
propagators in $V_{\textrm{1loop,top}}$ now contain different bulk 
masses for fields in the $Q$ and $U$ hypermultiplets.  Similarly, 
the $m_{\textrm{tree}}$ terms in $V_{\textrm{2loop}}$ differ for the 
left-handed and right-handed squarks.  The physical Higgs-boson mass 
can be calculated in the same manner as described in section 
\ref{sec:ewsb1} with the result shown in Fig.~\ref{fig:2DHiggs}.  

This shows that the previous discussion, with the condition 
$M_Q =M_U$, captures well the qualitative features of our model.
These include a correlation between a localized top quark and a heavy 
Higgs boson.  Specifically, increasing the degree of localization of 
either or both of the hypermultiplets causes the Higgs-boson mass to 
grow until the point of no EWSB is reached.  This is in contrast to the 
case of small $M_Q R$ and $M_U R$ where the Higgs-boson mass decreases 
as the bulk masses grow \cite{Barbieri:2002uk}.  In the case of the 
two Higgs theory, the potential is unstable in this region of 
parameter space.

\begin{figure}
\begin{center}
  \includegraphics[width=9cm]{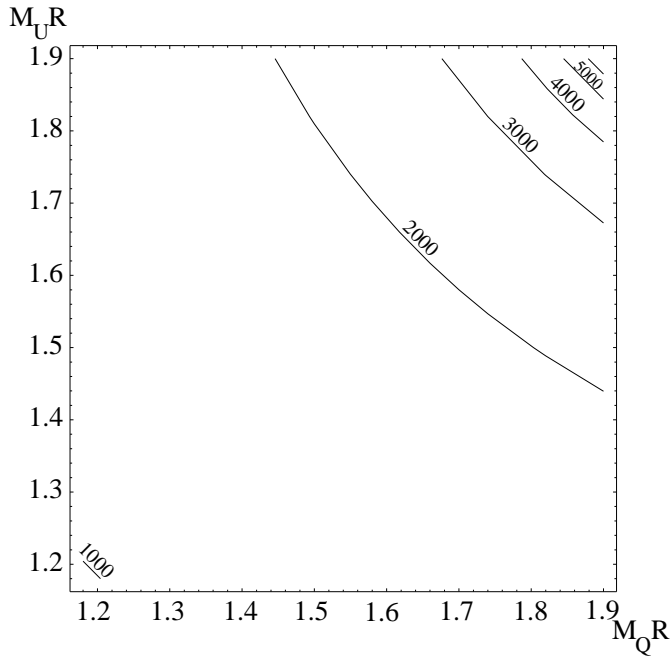}
\caption{Contours of $1/R$ in GeV as a function of $M_{Q}R$ and $M_{U}R$.}
\label{fig:2D1overR}
\end{center}
\end{figure}

Deviation from $M_Q =M_U$ also leaves unaltered the trend seen earlier 
that the localization of the top quark yields an increased $1/R$, as 
seen in Fig.~\ref{fig:2D1overR}.  This is again due to cancellation 
between the various contributions to the quadratic term in the Higgs 
potential.  Therefore, we find that essential features of the model 
do not change by deviating from the condition $M_Q = M_U$: as $M_{Q,U}$ 
are increased approaching to the point of no EWSB, $1/R$ increases 
due to the cancellation in the Higgs quadratic term, resulting in 
a heavier physical Higgs-boson mass through a larger correction to 
the Higgs quartic coupling coming from heavier stops.

\begin{figure}
\begin{center}
  \includegraphics[width=9cm]{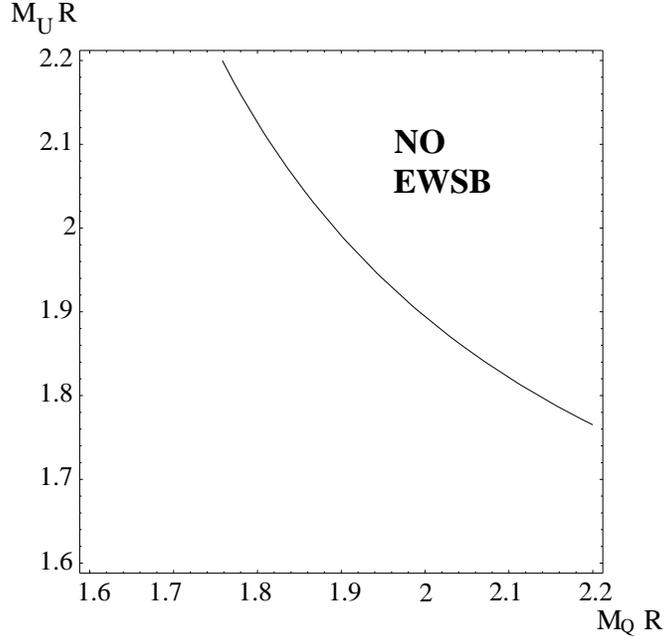}
\caption{Region of $\left(M_{Q}R,M_{U}R\right)$ space in which 
 electroweak symmetry is broken. Note that the ranges for $M_Q R$ 
 and $M_U R$ are different from those of Figs.~\ref{fig:2DHiggs}
 and \ref{fig:2D1overR}.}
\label{fig:2DEWSB}
\end{center}
\end{figure}

As either of the bulk masses, $M_Q$ or $M_U$, is decreased, 
$V^{(2)}_{\textrm{1loop,top}}$ becomes more negative; we do not have to 
reduce both $M_Q$ and $M_U$ to obtain sufficiently negative contribution 
from $V^{(2)}_{\textrm{1loop,top}}$. This opens up a new larger region 
of parameter space for which there is successful EWSB. 
In Fig.~\ref{fig:2DEWSB} we have shown the parameter region where 
successful EWSB occurs. This figure shows that Fig.~\ref{fig:2DHiggs} 
is somewhat misleading regarding the portion of parameter space that 
has been ruled out by direct Higgs searches.  While more than half 
of the parameter space $1.2 \lsim M_Q R, M_U R \lsim 1.9$ is ruled out 
yielding a too light Higgs boson, there is now an additional region that 
is experimentally viable in which either $M_Q$ or $M_U$ (but not both) 
is larger than $1.9/R$.

\begin{figure}
\begin{center}
  \includegraphics[width=14cm]{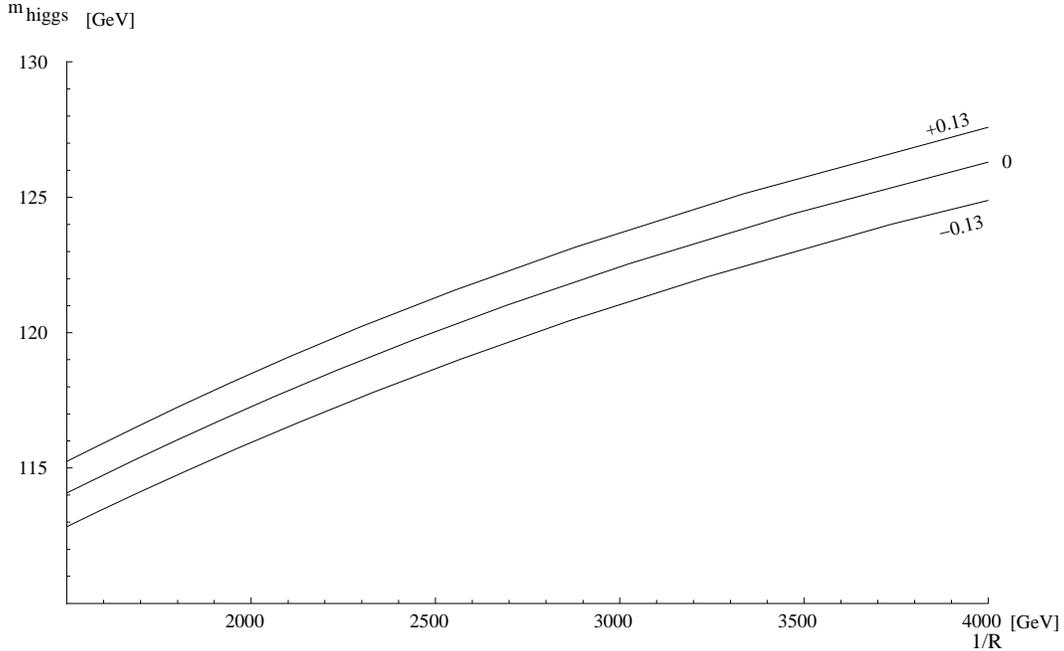}
\caption{The physical Higgs-boson mass as a function of $1/R$ for 
 three values of the Higgs tree-level mass squared, given in units 
 of $0.01/R^2$.}
\label{fig:Mhnonzero}
\end{center}
\end{figure}

Next we remove the restriction $M_{H_u}=0$ while maintaining 
$M_Q=M_U=M_t$.  The only significant effect of removing this restriction 
is to allow a tree-level mass for the Higgs doublet.  We consider only 
small tree-level mass squares (either positive or negative) so that 
the radiative effects are still dominant.  Specifically, we wish to
maintain the feature that as $M_t R$ approaches infinity, there is 
no EWSB.  Fig.~\ref{fig:Mhnonzero} demonstrates the case in which the 
magnitude of the tree-level mass squared is one half the sum 
$V^{(2)}_{\textrm{1loop,gauge}}+V^{(2)}_{\textrm{2loop}}$ in the 
localized limit.  Such a mass is of the same order of magnitude as 
that expected from a linearly divergent FI term, which corresponds 
to $25\%$ of the two-loop contribution in the localized limit. 
We find that the resulting effect is less than that arising from 
experimental uncertainty in the top quark mass.  Incidentally, 
increasing the tree-level mass with the positive (negative) sign
for $M_{H_u}$ increases (decrease) the region of $M_t R$ in which 
electroweak symmetry is broken.

Localization of the top quark by hypermultiplet masses has also been 
discussed in Ref.~\cite{Marti:2002ar}. They considered the limit of 
exact localization of $U$, $M_U R \rightarrow \infty$, and took a very 
high degree of localization of $Q$, $M_Q R = 2.6$, so that the $m_t/m_b$ 
ratio is entirely understood by the profile of $Q$. In this case they 
argue that EWSB is triggered by the two-loop top contribution, since 
the one-loop top term is negligible. However, our explicit two-loop 
calculation shows that their estimate of the two-loop contribution 
significantly exaggerates its effect, and that EWSB does not occur in 
this region.

\section{Complete Spectrum}
\label{sec:spectrum}

The value of $1/R$ determines in a simple way the spectrum of the
towers of gauginos, Higgsinos and gauge bosons: up to small EWSB
effects the lightest gauginos and Higgsinos are at $1/R$, whereas the 
first KK states of the vector towers are at $2/R$. The masses of the 
lightest scalars are of greater interest for experimental searches. 
For matter hypermultiplets with small hypermultiplet mass these are 
also at $1/R$, but as the hypermultiplet mass, $M$, increases the 
tree-level mass of the corresponding lightest scalar decreases, and 
vanishes exponentially for large $MR$. Hence in this limit the radiative 
and EWSB contributions to the mass become important. In the case of the 
theory with two Higgs hypermultiplets, there are additional scalars, 
$H^0$ and $H^+$ and their towers, from $H_d$. For $M_{H_d} = 0$, the 
scalars $H^0$ and $H^+$ are zero modes, acquiring mass from electroweak 
radiative corrections and from EWSB contributions. Possible radiative 
FI term contributions to scalar masses are taken to be included in $M$.

For any of these scalars, of charge $Q$ and hypercharge $Y$, we 
calculate the mass squared as
\begin{equation}
  m^2 = m^2_{\textrm{tree}} + m^2_{\textrm{rad}} + Y m_Z^2 - Q m_W^2,
\end{equation}
where $m_{\textrm{tree}}$ is the tree-level mass, dependent on the
mass of the hypermultiplet, $M$, to which the particle belongs and 
including the effect of the Yukawa coupling, $m_{\textrm{rad}}$ is the 
one-loop mass computed in localized approximation when $MR \gsim 1$, 
as in Eqs.~(\ref{eq:gauge1loophiggs}, \ref{eq:gauge1loophiggssu2}, 
\ref{eq:gauge1loophiggsu1}, \ref{eq:1loopmassQ}, \ref{eq:1loopmassU}), 
and finally the last terms arise from the $SU(2)$ and $U(1)$ $D$-terms 
after EWSB.

\begin{figure}
\begin{center}
  \includegraphics[width=15cm]{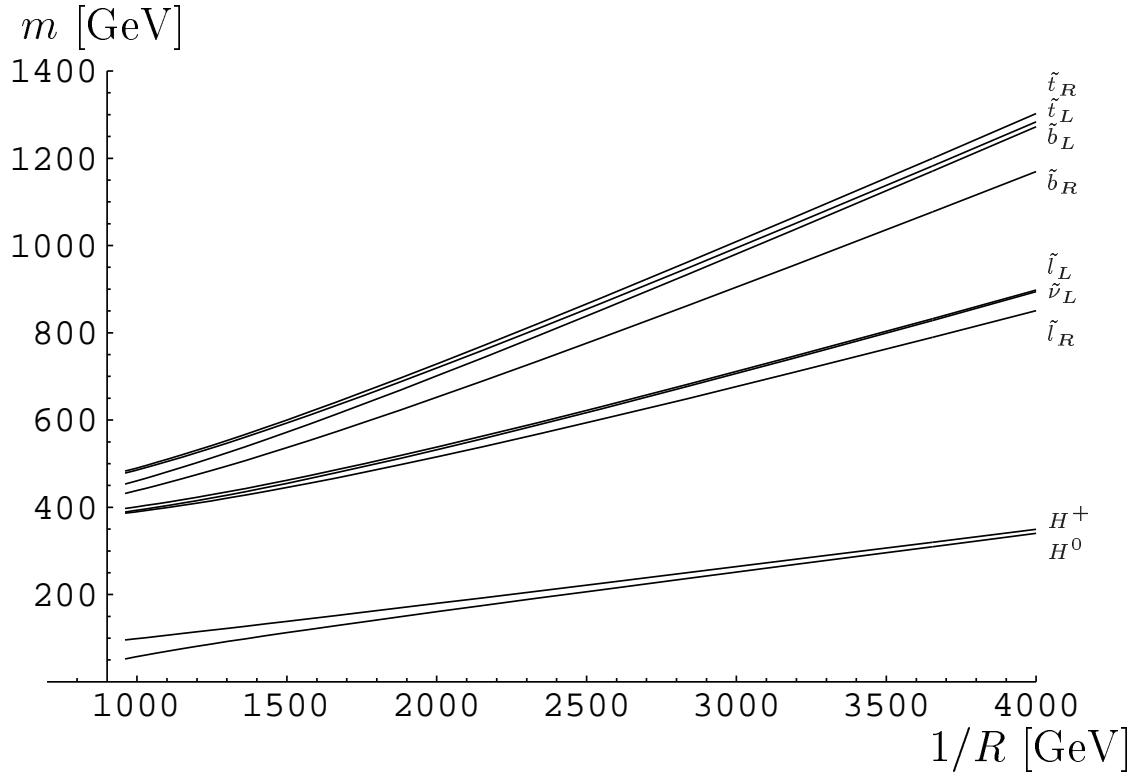}
\caption{Physical masses for the squarks and sleptons from hypermultiplets 
 with $M_Q = M_U = M_D = M_L = M_E = M$, and for the scalars of $H_d$ 
 with $M_{H_d} = 0$. As $1/R$ increases so does $M$, so that the squark 
 and slepton masses become dominated by the radiative contributions of 
 Eqs.~(\ref{eq:Qrad} -- \ref{eq:Erad}).} 
\label{fig:spectrum}
\end{center}
\end{figure}

The masses which are unequivocally determined are those belonging 
to the $Q$ and $U$ hypermultiplets that play a crucial role in EWSB, 
$\widetilde{t}_L, \widetilde{t}_R, \widetilde{b}_L$. They are given as 
functions of $1/R$ in Fig.~\ref{fig:spectrum} when $M_Q=M_U$. The 
radiative term dominates over all the other terms, hence a quasi-linear 
rise of the masses. A moderate variation in $M_U/M_Q$ has only a small 
influence on these masses, but a value larger than unity could reduce 
the mass of $\widetilde{t}_R$ relative to that of $\widetilde{t}_L, 
\widetilde{b}_L$.   

Fig.~\ref{fig:spectrum} shows masses for the other squarks and sleptons 
of the third generation in the case where all the matter hypermultiplet 
masses are taken equal to each other and $M_{H_d}=0$. For large values 
of $1/R$, these masses are dominated by the radiative contribution
\begin{eqnarray}
  m^2_{\textrm{rad}}(\widetilde{Q}) &=& \frac{0.060}{R^2},  \label{eq:Qrad}\\
  m^2_{\textrm{rad}}(\widetilde{U}) &=& \frac{0.064}{R^2},  \\
  m^2_{\textrm{rad}}(\widetilde{D}) &=& \frac{0.044}{R^2},  \\
  m^2_{\textrm{rad}}(\widetilde{L}) &=& \frac{0.0075}{R^2}, \\
  m^2_{\textrm{rad}}(\widetilde{E}) &=& \frac{0.0028}{R^2}, \label{eq:Erad}
\end{eqnarray}
computed here in the localized approximation, and including electroweak 
radiative corrections for the squarks as well as the sleptons. As any 
matter hypermultiplet mass is reduced, the corresponding scalar mass 
increases, asymptotically to $1/R$, due to the tree contribution. 
Only the matter with large hypermultiplet masses have light scalars.
The values given in Eqs.~(\ref{eq:Qrad} -- \ref{eq:Erad}) also apply to 
scalars of the first two generations if they originate in hypermultiplets 
with a large mass; the only difference is that the radiative top 
contribution of Eqs.~(\ref{eq:1loopmassQ}, \ref{eq:1loopmassU}) must 
be subtracted for $\widetilde{Q}$ and $\widetilde{U}$.

For the two Higgs hypermultiplet theory, $M_{H_{d}}=0$ ensures that 
the $H_d$-bosons are uniformly distributed in the bulk, so that 
the radiative masses are as in Eqs.~(\ref{eq:gauge1loophiggs} -- 
\ref{eq:gauge1loophiggsu1}). Note that, in this case, the mass of the 
neutral $H_d$ is below $100~{\rm GeV}$ for $1/R \lsim 1.4~{\rm TeV}$, 
but, since $H_d$ has no VEV this is not presently excluded.

\section{Phenomenology of Sparticle Production}
\label{sec:Pheno}

The precise phenomenology of sparticle production will depend upon the 
choice of the hypermultiplet masses. There are, however, a few features 
of this phenomenology that have a universal character. The lightest 
superpartner is a squark or slepton, most likely charged, which is 
stable or practically stable.\footnote{
Its instability can be due to a small $U(1)_R$-breaking effect or to 
its decay into a very light right-handed sneutrino. The latter case is 
motivated by having small Dirac neutrino masses accounted for by a 
Yukawa coupling at $y=0$ to a $N_R$ hypermultiplet strongly localized 
at $\pi R/2$. The neutrino mass is exponentially small, so that values 
of $M_{N_R} R$ need only be $2$--$4$ times larger than $M_{Q,U}$, 
depending on the localization of $L$.} 
This scalar is pair produced in a hadron collider, either directly or 
by cascade decay, via a strong interaction cross section determined 
by $1/R$.

If the squarks and sleptons of only the third generation are light 
enough to be relevant, there are three different cases to be considered:
\begin{itemize}
  \item The lightest sparticles are $\widetilde{t}_L, \widetilde{t}_R$ 
 and $\widetilde{b}_L$, as is the case where $M_Q=M_U=-M_{H_d}$ and all 
 other hypermultiplet masses are vanishingly small. In this case there 
 are two essentially degenerate super-hadrons $S^+$ and $S^0$, and
 their charge-conjugates, made of a squark, either $\widetilde{t}$ or 
 $\widetilde{b}$, depending on which is the lightest, and an antiquark, 
 which both appear as stable particles. The two other sparticles decay 
 into them by emission of soft hadrons. The masses of these sparticles 
 can be read off from Fig.~\ref{fig:spectrum}, as functions of $1/R$. 
 (The other superparticles shown in the figure, such as $\widetilde{b}_R$ 
 and the sleptons, are much heavier $\simeq 1/R$ in the present case.)
 The figure was drawn taking $M_D$ equal to $M_Q=M_U$, but the masses 
 for $\widetilde{t}_L, \widetilde{t}_R$ and $\widetilde{b}_L$ are 
 practically unchanged even when $M_D = 0$.

\item The lightest sparticle is $\widetilde{b}_R$, as is the case for 
 $M_Q=M_U=M_D$ with other hypermultiplet masses vanishingly small.
 The masses for the squarks in this case can be read off from 
 Fig.~\ref{fig:spectrum}. (The sleptons are much heavier of masses 
 $\simeq 1/R$).  Also in this case there are two degenerate super-hadrons 
 $S^+$ and $S^0$, produced either directly or by cascade with a
 larger cross section than in the previous case. Furthermore, the 
 heavier sparticles decay into them mostly in association with 
 $b$-quarks or $W$-bosons.

\item The lightest sparticle is a slepton, most likely charged, as 
 in the case when all hypermultiplet masses are equal (see 
 Fig.~\ref{fig:spectrum}). More precisely, in this case the lightest
 sparticle is a charged slepton, which can be pair produced by the
 Drell-Yan mechanism with an electroweak cross section or from the 
 cascade decay of the heavier states, always with at least a charged 
 lepton and, in most cases, with a $t$- or a $b$-quark. Note that 
 the lightest sparticle could also be a sneutrino, as in the case 
 $M_Q=M_U=-M_L$ and $M_E=0$.  A sneutrino would give rise to a missing 
 energy signal in association with $t$- or $b$-quarks from the 
 cascade decays.
\end{itemize}
Further light scalars could result from large hypermultiplet masses
for the first two generations.

\section{Bounds on $1/R$}
\label{sec:bounds}

The quasi-localization of some matter hypermultiplets gives rise to
new interactions which limit $1/R$ from below in a definitely stronger
way than in the case of matter homogeneously spread throughout the bulk, 
as in Ref.~\cite{Barbieri:2000vh}. The strength of these interactions 
critically depends on the localization of the first two generations, 
i.e. on the hypermultiplet masses $M_{1,2}$, and also depends in general 
on the gauge quantum numbers inside one generation.

Correspondingly, the strongest limits arise in two cases:
\begin{enumerate}
\def\labelenumi{(\theenumi)}
\item when the first generation is mostly localized, from 4-fermion 
 operators generated by exchanges of KK gauge bosons.
\item when the first two generations have their hypermultiplet masses 
  different from each other or from the one of the third generation, 
  from their Flavor Changing Neutral Current (FCNC) effects.
\end{enumerate}

In view of this, for definiteness, we consider two different cases:
\begin{enumerate}
\def\theenumi{\roman{enumi}}
\def\labelenumi{(\theenumi)}
\item $M_1=M_2=M_3$ for the entire family multiplets, 
\item $M_1=M_2=0$, with $M_3$ as in section \ref{sec:ewsb1}.
\end{enumerate}

When all masses are equal, the case (i), there is no new FCNC effect. 
On the contrary, a new significant effect occurs through the couplings 
of the KK $W$-bosons, of masses $2n/R$, to the first generation 
lepton-doublet $L_1$, described by the effective Lagrangian
\begin{equation}
  {\mathcal L}_{\rm eff} 
    = \frac{A}{16} g^2 R^2 (L_1 \gamma_{\mu} \tau^a L_1)^2.
\end{equation}
Here, $A$ is a normalization factor depending upon $M_1 R$ through the 
zero-mode wavefunctions, which vanishes for $M_1 R=0$ and is close to 
unity for $M_1 R \simeq 1$--$2$. In order not to disturb the success 
of EWPT fit, for $M_1 R = 1$--$2$ a bound on $1/R$ of $1.4~{\rm TeV}$ 
arises \cite{Strumia:1999jm}.

Suppose now that we consider the case where the first two generations, 
unlike the third one, are uniformly spread in the bulk ($M_1=M_2=0$), 
so that this effect is absent. In this case it is the difference 
between the couplings of the KK gluons to the first two generations 
and to the third one that gives the largest effect, as calculated in 
Ref.~\cite{Delgado:1999sv}. If one assumes mixing angles and phases of 
the down-quark Yukawa-coupling matrix comparable to those of the 
Cabibbo-Kobayashi-Maskawa matrix, the strongest bound arises from 
$CP$-violating $\epsilon$-parameter in $K$ physics and is about 
$1/R \gtrsim 2~{\rm TeV}$.

From these considerations we conclude that the range of values for 
$1/R$ compatible with EWSB, $1/R = 1.5$--$3.5~{\rm TeV}$, could 
give rise to some interesting indirect effects either in EWPT or 
in flavor physics.

\section{Conclusions}

We have constructed a theory of ElectroWeak Symmetry Breaking (EWSB) 
with supersymmetry broken by boundary conditions in a fifth dimension, 
which is determined to have a scale $1/R \simeq 1.5$--$3.5~{\rm TeV}$. 
The only particles beyond those of the standard model which must 
be lighter than $1/R$ are the three squarks $\widetilde{t}_L, 
\widetilde{b}_L$ and $\widetilde{t}_R$, which have masses approximately 
proportional to $1/R$ and in the range $500$--$1200~{\rm GeV}$.

In Table~\ref{table:comp} we compare this theory with the Standard 
Model (SM), the Minimal Supersymmetric Standard Model (MSSM) and the 
Constrained Standard Model (CSM).  By the constrained standard model, 
we mean the theory introduced in Ref.~\cite{Barbieri:2000vh} 
together with the possibility of small hypermultiplet masses 
\cite{Barbieri:2001cz, Barbieri:2002uk}. The only crucial difference 
of the model in the present paper is that the top quark has a large 
hypermultiplet mass, causing it to be approximately localized at a 
boundary of the fifth dimension. The first row of Table~\ref{table:comp} 
shows whether each model, considered as an effective field theory below 
some cutoff scale $\Lambda$, provides a physical theory of EWSB. Can 
the electroweak scale, $v$,  be computed in terms of some parameters of 
the effective theory in a way which is relatively insensitive to physics 
at the cutoff and therefore to $\Lambda$? This of course is the great 
failing of the SM, motivating the introduction of the other models. 
The second row shows whether any of these models predicts new particles 
which would have already been discovered by direct searches. Strictly 
speaking none does, but in the case of the MSSM this is because the 
parameter space of the model allows cancellations so that the 
superpartners can be made unnaturally heavy. The third row shows that 
none of the theories is in conflict with ElectroWeak Precision Tests
(EWPT), although excessive contributions to the $\rho$ parameter might
have been expected in the CSM, as it has a low cutoff scale, $\Lambda 
\approx 2~{\rm TeV}$. High scale gauge coupling unification is only 
possible in the SM and MSSM, where $\Lambda$ is above the unification 
scale, and is successful only for the MSSM, as shown in the fourth row. 
\begin{table}
\begin{center}
\begin{tabular}{|c||c|c|c|c|} \hline
                  & SM  & MSSM & CSM & This paper \\ \hline \hline
EWSB              & $-$ & $+$  & $+$ &    $+$     \\ \hline
Direct searches   & $+$ & $?$  & $+$ &    $+$     \\ \hline
EWPT              & $+$ & $+$  & $?$ &    $+$     \\ \hline
Gauge coup. unif. & $-$ & $+$  & $-$ &    $-$     \\ \hline
\end{tabular}
\caption{A comparison of models for EWSB.}
\label{table:comp}
\end{center}
\end{table}

The MSSM has a very plausible physical origin for EWSB: a negative 
Higgs mass-squared, $m_\phi^2$, induced radiatively by the top-quark 
Yukawa coupling. However, the top Yukawa coupling is large and, even 
though it is radiative, this effect is very powerful
\begin{equation}
  m_\phi^2 \approx 
    - {\ln(\Lambda / \widetilde{m}) \over 30} \widetilde{m}^2,
\label{eq:m^2susy}
\end{equation}
partly because of the large logarithm. Consequently, the scale 
$\widetilde{m}$ of colored superpartners is expected to be close to 
$v$, and this is especially problematic when the supersymmetry breaking 
leads to non-colored superpartners significantly lighter than the 
colored ones. In the CSM the calculation is extended to include the 
KK modes of the top quark, with the result that the Higgs mass squared 
is finite and independent of $\Lambda$
\begin{equation}
  m_\phi^2 \approx 
    -9 \left( {0.01 \over R^2} \right).
\label{eq:m^2csm}
\end{equation}
The masses for both colored and non-colored superpartners are predicted
to be at $1/R \simeq 400~{\rm GeV}$; the issue of fine-tuning does not
arise, and these superpartners could be readily discovered or excluded. 
However, such low values of $1/R$, and therefore $\Lambda$, might have 
been discovered at LEP in EWPT.

The quasi-localized top quark studied in this paper has the virtue 
of exponentially suppressing any one-loop top contribution, such as 
Eqs.~(\ref{eq:m^2susy}, \ref{eq:m^2csm}). Relevant contributions to 
the Higgs mass squared are then provided by electroweak gauge 
interactions at one loop:
\begin{equation}
  m_\phi^2 \simeq 
    0.76 \left( {0.01 \over R^2} \right),
\label{eq:m^21loopg}
\end{equation}
and two-loop diagrams involving the top-quark Yukawa interaction
\begin{equation}
  m_\phi^2 \simeq 
    -0.49 \left( {0.01 \over R^2} \right).
\label{eq:m^22loopt}
\end{equation}
The sum of these contributions is $30$ times smaller than the 
unsuppressed one-loop top contribution of Eq.~(\ref{eq:m^2csm}), and
therefore leads to an increase in $1/R$ from the CSM value by about 
a factor of $6$ to the region of $2.5~{\rm TeV}$. However, this sum 
is positive: an exactly localized top quark does not lead to any EWSB. 
This means that the localization must not be complete: the one-loop 
top contribution must not be negligible. At first sight this looks 
like another fine tune, but it is not: the one-loop top contribution 
is suppressed by a factor $\exp(-\pi M_t R)$ which is in the desired 
range of $10^{-1}$--$10^{-2}$ for $M_t R \approx 1$--$2$.

\section*{Acknowledgements}

Y.N. thanks the Miller Institute for Basic Research in Science 
for financial support.  The work of L.H., Y.N., T.O. and S.O 
was supported in part by the Director, Office of Science, Office of 
High Energy and Nuclear Physics, of the U.S. Department of Energy 
under Contract DE-AC03-76SF00098, and in part by the National Science 
Foundation under grant PHY-00-98840. The work of R.B., G.M. and M.P. 
has been partially supported by MIUR and by the EU under TMR contract 
HPRN-CT-2000-00148.  This material is based upon work supported under 
a National Science Foundation Graduate Research Fellowship.

\section*{Appendix A}

In this appendix, we discuss how many bulk mass parameters the theory 
possesses and how many of them enter the EWSB calculation.  Without loss 
of generality the bulk mass matrix for each charge sector can be taken 
diagonal, so that there is a separate mass parameter for each matter 
and Higgs hypermultiplet.  The masses for the scalar and fermion 
components take the form
\begin{eqnarray}
  \mathcal{L}_m &=& \cdots -\psi^c\partial_y\psi 
                    - M \eta(y) \left(\psi^c\psi + \mathrm{h.c.}\right)
    - M^2 \left(|\phi|^2 +|\phi^c|^2 \right) 
\nonumber \\
  && +2M\left( \delta(y) +\delta(y-\pi R/2)\right)
    \left(|\phi|^2 - |\phi^c|^2 \right), 
\end{eqnarray}
where $\eta(y) = +1$ ($-1$) for $y > 0$ $(< 0)$, and the $\partial_y$ 
piece is included because what matters is a relative sign between 
$\partial_y$ and $M$.  These mass terms have 
a brane contribution to maintain the form of the unbroken local 
supersymmetry \cite{Barbieri:2002uk}. Despite the presence of so many 
parameters, the physics of EWSB is sensitive to only three of them: 
$M_{Q_3}$, $M_{U_3}$ and $M_{H_u}$. It is perhaps obvious that the masses 
for the lighter generation quarks are irrelevant --- they have small 
Yukawa couplings which give only small radiative contributions to the 
scalar potential --- but it is not obvious that $M_{D_3}$ is irrelevant. 
A large value for $M_{Q_3}$ localizes $b_L$ largely on the brane distant 
from the bottom Yukawa coupling, so that the 5D bottom Yukawa coupling 
must be large to overcome the wavefunction suppression. Nevertheless, 
we find that the radiative contribution to the Higgs potential through 
the bottom Yukawa coupling is always suppressed by $(m_b/m_t)^2$.

We consider here only the case of a single Higgs hypermultiplet.
Everything that follows may be directly generalized to the two Higgs
case for moderate value of $M_Q R$.  The relevant part of the Lagrangian 
for studying the contribution of the radiative correction from the 
bottom quark Yukawa coupling to the scalar potential is:
\begin{equation}
  \mathcal{L} 
    = \lambda_{t} \delta(y) (\widetilde{q}F_{U}h 
      + F_{Q}\widetilde{t}h - qth) 
    - \lambda_{b} \delta(y-\pi R/2) (\widetilde{q}^{c*}F'_{D}h^{*}
      + F'_{Q}\widetilde{b}^{c*}h^{*} - qbh^{*}).
\label{eq:app-lag}
\end{equation}
Here, the chiral supermultiplets under the $N=1$ supersymmetry acting 
at $y=0$ are given by
\begin{eqnarray}
  H     &=& (h,\widetilde{h},F_{H}), \\
  Q_{3} &=& (\widetilde{q},q,F_{Q}), \\
  U_{3} &=& (\widetilde{t},t,F_{U}).
\end{eqnarray}
and the chiral supermultiplets under the $N=1$ supersymmetry acting 
at $y=\pi R/2$ are given by
\begin{eqnarray}
  H^{\prime c} &=& (-h^{*},\widetilde{h}^{c},F'_{H}), \\
  Q'_{3}       &=& (\widetilde{q}^{c*},q,F'_{Q}), \\
  D'_{3}       &=& (\widetilde{b}^{c*},b,F'_{D}).
\end{eqnarray}
The Lagrangian in Eq.~(\ref{eq:app-lag}) can be derived from the 
superpotential term $W = \lambda_{t} \delta(y) (Q_{3}U_{3}H) + 
\lambda_{b} \delta(y-\pi R/2) (Q'_{3}D'_{3}H^{\prime c})$.

In terms of mixed momentum-position propagators, the ratio of bottom 
to top Yukawa contributions in the Higgs mass squared clearly depends 
on the ratios $G_q(k_4; 0, 0) / G_q(k_4; \pi R/2, \pi R/2)$, 
$G_{\widetilde{q}}(k_4; 0, 0) / G_{\widetilde{q}^{c*}}(k_4; \pi R/2, 
\pi R/2)$ and $G_{F_{Q}}(k_4; 0, 0) / G_{F'_{Q}}(k_4; \pi R/2, 
\pi R/2)$. These ratios of propagators are all equal in the infrared, 
which dominates the loop integral, and given by $\exp(-\pi M_{Q} R)$. 
This exactly cancels the enhancement of the 5D bottom Yukawa coupling 
due to the small wavefunction overlap.  Therefore, the contribution 
to the Higgs mass squared due to the bottom Yukawa interaction is 
down by a factor of $(m_b/m_t)^2$ and can be safely neglected. We can 
similarly neglect all the other Yukawa contributions relative to the 
top one. The most general such theory of EWSB is therefore 
parameterized by a three dimensional space spanned by 
$(M_{Q_{3}},M_{U_{3}},M_{H_{u}})$.

\section*{Appendix B}

In this appendix we list propagators $G_i(p;0,0)$ for various 
components of a matter hypermultiplet with a bulk mass $M$:
\begin{eqnarray}
  G_{\varphi}(p;0,0) 
    &=& \frac{\sinh[\sqrt{p^2+M^2}\pi R/2]}
    {\sqrt{p^2+M^2}\cosh[\sqrt{p^2+M^2}\pi R/2] 
    - M\sinh[\sqrt{p^2+M^2}\pi R/2]},
\\
  G_{\psi}(p;0,0) 
    &=& \frac{\sqrt{p^2+M^2}}{\spur{p}} \coth[\sqrt{p^2+M^2}\pi R/2] 
    + \frac{M}{\spur{p}},
\\
  G_{F}(p;0,0) 
    &=& \frac{\cosh[\sqrt{p^2+M^2}\pi R/2]
    + \frac{M}{\sqrt{p^2+M^2}}\left(1+\frac{p^2}{2M^2}\right)
      \sinh[\sqrt{p^2+M^2}\pi R/2]} 
    {\frac{1}{2M}\cosh[\sqrt{p^2+M^2}\pi R/2]
    + \frac{1}{2\sqrt{p^2+M^2}}\sinh[\sqrt{p^2+M^2}\pi R/2]},
\end{eqnarray}
where $\varphi$, $\psi$ and $F$ represent the scalar, fermion and 
auxiliary field components, respectively; $p$ is an Euclidean momentum.

\end{document}